# Inverse Design of Perfectly Transmitting Eigenchannels in Scattering Media

M. Koirala,[1] R. Sarma,[2] H. Cao,[2] and A. Yamilov[1, *]

[1]*Department of Physics, Missouri University of Science and Technology, Rolla, Missouri 65409,USA*
[2]*Department of Applied Physics, Yale University, New Haven, CT, 06520, USA*
(Dated: March 14, 2017)

Light-matter interactions inside turbid medium can be controlled by tailoring the spatial distribution of energy density throughout the system. Wavefront shaping allows selective coupling of incident light to different transmission eigenchannels, producing dramatically different spatial intensity profiles. In contrast to the density of transmission eigenvalues that is dictated by the universal bimodal distribution, the spatial structures of the eigenchannels are not universal and depend on the confinement geometry of the system. Here, we develop and verify a model for the transmission eigenchannel with the corresponding eigenvalue close to unity. By projecting the original problem of two-dimensional diffusion in a homogeneous scattering medium onto a one-dimensional inhomogeneous diffusion, we obtain an analytical expression relating the intensity profile to the shape of the confining waveguide. Inverting this relationship enables the inverse design of the waveguide shape to achieve the desired energy distribution for the perfectly transmitting eigenchannel. Our approach also allows to predict the intensity profile of such channel in a disordered slab with open boundaries, pointing to the possibility of controllable delivery of light to different depths with local illumination.

Interference of scattered waves in random media gives rise to well-known phenomena such as enhanced backscattering, Anderson localization and universal conductance fluctuation. These phenomena are general and occur not only for electromagnetic waves, but also for acoustic, electronic and other kinds of waves [1, 2]. Recently, there has been a growing interest in another interference effect – formation of perfectly transmitting channels [3, 4], which can greatly enhance the total transmission through opaque media [5–8]. In addition, the perfectly transmitting channels have energy density buildup deep inside the medium [7, 9–11], opening the possibility of enhancing linear and non-linear light-matter interactions inside turbid media. Recent advances of optical wavefront shaping techniques [12–16] enabled direct coupling of incident light to perfectly transmitting channels [11], making the depth profile of energy density dramatically different from the typical decay in a diffusive medium. To unlock the full potential of this approach for tailoring light-matter interactions in turbid media, it becomes imperative to understand what determines the spatial structure of the perfectly transmitting channels.

Recently two theoretical models have been put forward to describe the spatial profile of the perfectly transmitting channels in lossless diffusive media. Davy et al [9] applied the supersymmetry theory to wave propagation in a quasi-one-dimensional random system and related the intensity profile to the return probability (RP) of diffusive waves. Ojambati et all [10, 17] proposed that the perfectly transmitting channel in a disordered slab is related to the fundamental mode (FM) of the one-dimensional (1D) diffusion equation. Although both models predict correctly the depth of the maximum energy density [7, 9, 11, 18], they disagree quantitatively in terms of the depth profile for the perfectly transmitting channel. So far, both models have been applied only to one-dimensional diffusion.

We have studied light transport in quasi-two-dimensional disordered systems, and showed that the spatial structure of transmission eigenchannels can be modified by the confinement geometry [19]. For example, by adjusting the shape of the reflecting boundary of a disordered waveguide, the depth at which the energy density of a high transmission channel reaches the maximum can be moved. This enables enhancing light-matter interaction at different location inside the random medium. For many applications, inverse design is needed, namely, to design the confinement geometry to achieve the desired depth profile of energy density inside a diffusive system. This requires a prior knowledge of the relation between the geometry of the system and the spatial structure of the transmission channels. However, there is currently no theoretical model capable of establishing such relation.

In this work, we consider a two-dimensional disordered waveguide with an *arbitrary* shape, and develop a theoretical model to predict the spatial structure of the perfectly transmitting eigenchannel in the regime of diffusive transport. We further employ a projection technique, developed in physical chemistry for the particle diffusion in confined geometries [20], to obtain an analytical relation between the depth profile of the perfectly transmitting eigenchannel and the geometry of the waveguide. With this relation, we perform the inverse design of the waveguide shape to realize the desired energy distribution for the perfectly transmitting eigenchannel. Finally, we predict the depth profile for the perfectly transmitting eigenchannel in an open slab geometry with local illumination, by simulating the lateral beam spread in a waveguide of expanding width. We find that the depth of the maximum of intensity increases with the size of the impinging beam, which provides an insight into controlling the energy distribution inside a diffusive slab with local illumination.

The transmission matrix, which connects the transmitted fields to the incident fields, contains the complete information about wave transport through the disordered system. Transmission eigenchannels are introduced via singular value decomposition of the transmission matrix $\hat{t} = \hat{U}\hat{\tau}^{1/2}\hat{V}^\dagger$. Here $\hat{\tau}$ is the diagonal matrix of eigenvalues of $\hat{t}^\dagger \hat{t}$ that represent the transmittance of each eigenchannel; $\hat{V}$ is a unitary matrix that maps the incoming modes onto the eigenchannels; and $\hat{U}$ is a unitary matrix that maps the eigenchannels onto the outgoing modes. In the regime of diffusive transport, the transmission eigenvalues have a universal bimodal distribution, independent of both the microscopical details of the disorder and the boundary shape of the system. It consists of two peaks at $\tau \sim 0$ and $\tau \sim 1$, which correspond respectively to closed and open eigenchannels [21–24].

Here, we consider a two-dimensional (2D) waveguide with reflecting boundaries at $y = \pm W(z)/2$. The region $0 \leq z \leq L$ is filled with lossless scattering medium characterized by the transport mean free path $\ell \ll L$. The waveguide width $W(z)$ can be either larger or smaller than the length $L$, corresponding to slab or quasi-1D geometry. Our aim is to predict the depth profile (cross-section averaged intensity) of the perfectly transmitting eigenchannel (PTE) with $\tau \simeq 1$, $\phi(z) \equiv W^{-1}(z) \int_{-W(z)/2}^{W(z)/2} \langle I_{PTE}(\mathbf{r})\rangle dy$, where $I_{PTE}(\mathbf{r})$ is the intensity and $\langle...\rangle$ denotes the ensemble averaging. $I_{PTE}(\mathbf{r}) \equiv |E_{PTE}(\mathbf{r})|^2$, and $E_{PTE}(\mathbf{r})$ is the solution of the wave equation with the incident wave given by the eigenvector (a column vector in $\hat{V}$) corresponding to the eigenvalue $\tau \simeq 1$.

Using the diagrammatic theory [9] we establish the connection between the spatial profile of PTE and the return probability of diffusive waves. We first consider a rectangle-shaped waveguide of constant width $W < L$, then generalize the result to an arbitrary shaped system. We propose a phenomenological model based on the return probability $\tilde{G}(z; z)$ of light to a cross-section of the waveguide at depth $z$, which gives the depth profile of PTE $\phi(z) = \tilde{G}(z; z)$. $\tilde{G}(z; z') \equiv W^{-2}(z) \int \int G(y, z; y', z') dy dy'$, and $G(\mathbf{r}; \mathbf{r}')$ is the solution of the 2D diffusion equation with the additional self-action term on the right-hand-side of the following equation:

$$-\nabla_\mathbf{r}^2 G(\mathbf{r}; \mathbf{r}') = [1 + \alpha G(\mathbf{r}; \mathbf{r}')]\delta(z - z')/W(z). \quad (1)$$

The self-action term accounts for the effect of interference of waves that return after multiple scattering. It reflects the fact that upon the return, the coherent sum of the fields leads to the cross terms in the total intensity, similar to the weak localization correction [25]. With the proper choice of $\alpha$ (to be determined below), this equation can be solved inside the disordered waveguide ($0 \leq z \leq L$, $|y| \leq W(z)/2$) with the open boundary conditions at the two ends ($z = 0, L$), $[z_0 \partial G(\mathbf{r}; \mathbf{r}')/\partial z \mp G(\mathbf{r}; \mathbf{r}')]_{z=0,L} = 0$, where $z_0 = (\pi/4)\ell$ is the 2D extrapolation length [1]. The solution can be readily obtained numerically.

To obtain the analytical solution, we employ a projection technique that was developed in the study of diffusion of electrolytes in nano-pores [20]. This technique reduces the process of solving 2D diffusion equation in a complex geometry to a solution of the one-dimensional diffusion equation (along $z$-axis) with an effective diffusion coefficient that varies with $z$:

$$\mathcal{D}(z) = W(z)/\left[1 + (W'(z)/2)^2\right]^{1/3}. \quad (2)$$

Eq. (1) is then transformed to yield $\tilde{G}(z; z')$ directly

$$-\frac{\partial}{\partial z}\mathcal{D}(z)\frac{\partial}{\partial z}\tilde{G}(z; z') = \left[1 + \alpha \tilde{G}(z, z')\right]\delta(z - z'), \quad (3)$$

while the boundary conditions at $z = 0, L$ remain the same. This method suits our problem because we are interested in the depth dependence of the cross-section-averaged intensity profile.

We stress that the $z$-dependent diffusivity $\mathcal{D}(z)$ arises from the varying width $W(z)$ in a purely diffusive waveguide where the localization corrections are negligible. In the regime where localization corrections are significant, the projection ansatz used to obtain Eq. (3) from Eq. (1) is still applicable with the effective diffusion coefficient $\mathcal{D}(z) \times D(z)/D_0$, where $D(z)$ is the cross-section averaged value of the position-dependent diffusion coefficient [26–28] due to the localization-induced renormalization and $D_0$ is its unrenormalized value.

Solution of Eq. (3) can be obtained in the closed form

$$\phi(z) = \frac{(1 - \tilde{\alpha})F(z)}{[1 - \tilde{\alpha}F(z)]} \quad (4)$$

$$F(z) = \frac{4\left(\frac{z_0}{\mathcal{D}(0)} + \int_0^z \frac{dz'}{\mathcal{D}(z')}\right)\left(\frac{z_0}{\mathcal{D}(L)} + \int_z^L \frac{dz'}{\mathcal{D}(z')}\right)}{\left(\frac{z_0}{\mathcal{D}(0)} + \frac{z_0}{\mathcal{D}(L)} + \int_0^L \frac{dz'}{\mathcal{D}(z')}\right)^2}$$

$$\tilde{\alpha} = \frac{\alpha}{4}\left(\frac{z_0}{\mathcal{D}(0)} + \frac{z_0}{\mathcal{D}(L)} + \int_0^L \frac{dz'}{\mathcal{D}(z')}\right). \quad (5)$$

$\phi(z)$ has been normalized so that $\max[\phi(z)] = 1$ and $F(z)$ is an auxiliary function, which corresponds to the normalized solution of the same set of equations with $\alpha = 0$. The value of $\alpha$ (or $\tilde{\alpha}$) can be found from the waveguide with constant width.

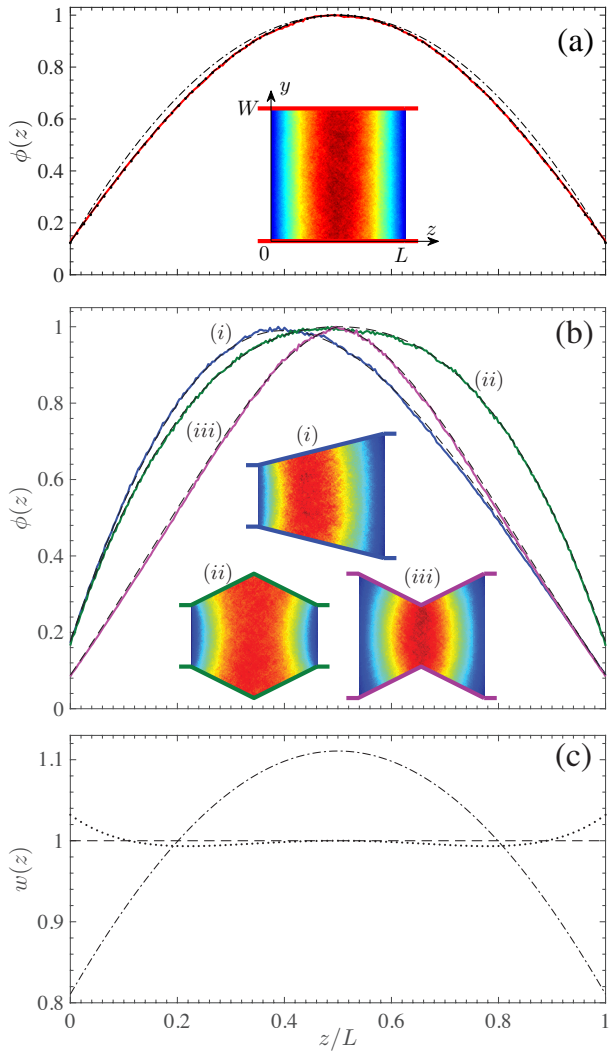

FIG. 1. (Color online) Comparison of previous models to ours in predicting the perfectly transmitting eigenchannels (PTEs) in the diffusive waveguides. (a) The cross-section integrated intensity $\phi(z)$ of the PTE in a constant width waveguide, computed numerically (bold solid line) and predicted by the RP-model (dash-dotted line), the FM-model (dotted line), and our model Eq. (4) (dashed line). The inset is 2D intensity distribution of the PTE, computed numerically, throughout this waveguide with $L/\ell = W/\ell \simeq 18.3$. (b) $\phi(z)$ for the PTEs in three waveguides of varying cross-section. Blue color: expanding waveguide, green color: lantern waveguide, purple color: bow-tie waveguide. Solid lines: numerical simulation, dashed lines: our model Eq. (4). The inset are the numerically calculated 2D intensity distribution for the PTEs in the three waveguides. The waveguide length and width at the widest point are identical to those of the waveguide shown in (a). The width at the narrowest point is equal to a half of that at the widest point. (c) The waveguide width $w_0(z)$ obtained by inverting (via Eq. (6)) the depth profile $\phi_0(z)$ predicted in the rectangle geometry by the RP-model (dash-dotted line), the FM-model (dotted line), and our model Eq. (4) (dashed line). $z_0 \ll L$ is assumed for clarity. The FM model shows less deviation from a constant width than the RP-model, but only our model produces the consistent result of constant width.

To test the analytical solution, we compare it to the numerical solution obtained by directly solving the wave equation with KWANT simulation package [29, 30]. Figure 1a compares the profile of PTE computed numerically to the predictions of the two previously developed models [9, 17]. Although RP-model deviates from the numerical solution, both models agree well at $z = 0, L$. They give $\phi^{(RP)}(0) = (8/\pi) \times (\ell/L) \simeq 2.546 \times (\ell/L)$ and $\phi^{(FM)}(0) = (\pi^2/4) \times (\ell/L) \simeq 2.467 \times (\ell/L)$. The knowledge of $\phi(z)$ at one point is enough to recover the value of the coefficient $\tilde{\alpha}$. Taking the former value, we find $\tilde{\alpha} = 1 - 8/\pi^2 \simeq 0.189$. Importantly, we observe that $\tilde{\alpha}$ is a non-zero numerical constant independent of system parameters, such as the transport mean free path and the system dimension, underlining the universality of the self-action term on the right-hand-side of Eqs. (1,3). The result of Eqs. (4) is shown as dashed line in Fig. 1a, it agrees well with the result of numerical simulations. Furthermore, in Fig. 1b we verify that the same value of $\tilde{\alpha}$ applies to waveguides with varying cross-section. We stress that the agreement between numerical simulations and our model is achieved with no fitting parameters.

The closed-form analytical solution given in Eq. (4) establishes the relation between the shape of the diffusive waveguide and the depth profile of the PTE, thus enabling the inverse design. By introducing a normalized width function $w(z) = [W(z)/L] \times \int_0^L dz'/W(z')$, the dimensionless conductance $g = k\ell / \left[2 \int_0^L dz'/W(z')\right]$, and neglecting the extrapolation length $z_0$, we obtain an expression for waveguide boundary function $w_0(z)$ in terms of the depth profile $\phi_0(z)$

$$w_0(z)/\left[1 + \left(\frac{g}{k\ell}w_0'(z)\right)^2\right]^{1/3} = \qquad (6)$$
$$\left[(1-\tilde{\alpha})(1-\tilde{\alpha}(1-\phi_0(z)))^3(1-\phi_0(z))/\phi_0'^2(z)\right]^{1/2},$$

where $k = 2\pi/\lambda$ is wave number and $\phi_0(z)$ satisfies the boundary condition $\phi_0(z = 0, L) = 0$. The extrapolation length at the boundary $z = 0, L$ can be accounted for by the following scaling

$$\phi_0(z) \to \phi(z) = \phi_0\left((z+z_0)/(L+2z_0)\right) \qquad (7)$$
$$w_0(z) \to w(z) = w_0\left((z+z_0)/(L+2z_0)\right). \qquad (8)$$

In the SI [30], we present a table of $\phi(z)$ for the waveguide geometries $w(z)$ shown in Fig. 1b.

The relation between $w(z)$ and $\phi(z)$ allows us to infer the shape of the waveguide from the depth profile of the PTE. For the depth profile of PTE predicted by the RP- and FM-models for the rectangle waveguide, c.f. Fig. 1c, we derive the corresponding waveguide shape, as shown in Figs. 1c. The shape predicted by the RP model corresponds to a waveguide with the width variation of up to 20%. In contrast, FM-model is more accurate. However, in other waveguide geometries, as shown in the SI [30], the PTE profiles predicted by the FM-model are inconsistent with the results of the numerical simulations.

To demonstrate the power of inverse design, we change the universal profile of PTE in constant-width waveguides to a highly unusual profile of a triangle. According to Eq. (6), to have $\phi_0(z)$ increase linearly with $z$, we find the waveguide width $w(z) \propto \sqrt{z}$ in the leading order of $\tilde{\alpha}$. This allows us to design waveguides that support PTE with a triangular profile, as shown in Fig. 2. The waveguide boundary is described by $W(z) = W_c + \Delta W \sqrt{|z - z_c|/z_c}$, where $z_c$ denotes the depth at which the width is the narrowest. It also closely corresponds to the maximum of the intensity profile of the PTE. The results in Fig. 2, obtained for two different values of $z_c/L = 1/2$ and $2/3$, show that it is possible to obtain waveguide geometries that have maximum concentration of energy at the desired depth.

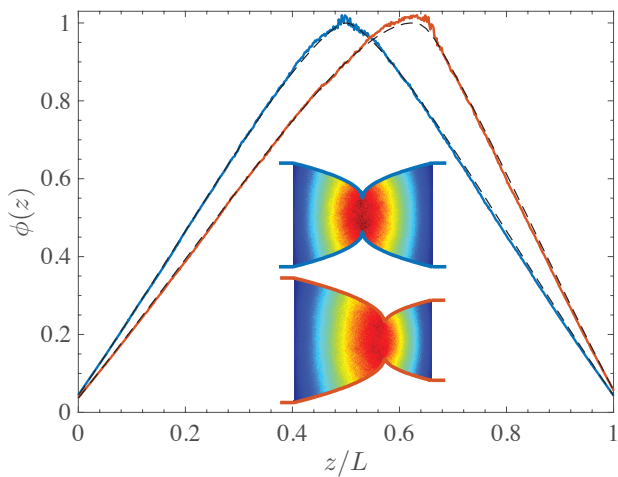

FIG. 2. (Color online) Design of waveguide shape to achieve the desired depth profile of PTE. The cross-section averaged intensity $\phi(z)$ showing an abnormal triangular dependence on the depth is obtained by the inverse design of the waveguide width $w(z)$ by Eq. (6). Bold solid lines are the result of numerical simulation, and dashed lines are the prediction by our model. The two waveguides, shown in the inset, have the PTE intensity maximum at the depth $z_c/L = 1/2$ (blue line) and $2/3$ (red line). The 2D intensity distribution of PTE are plotted in insets.

Finally, the inverse design introduced above provides an insight to controlling the depth profile of light intensity inside *a disordered slab* with local illumination. Such a geometry is common in optical experiments, and it is different from the waveguide geometry because the light will diffuse laterally as it penetrates deeper into the slab. A waveguide expanding linearly at 45° angle can be used as a proxy for studying the lateral diffusion in the slab geometry with local illumination [31]. In particular we consider a waveguide with expanding width $W(z) = W_1 + (W_2 - W_1) \times (z/L)$, and the tapering angle of the waveguide boundary is $\theta = \arctan[(W_2 - W_1)/2L] = 45°$, to mimic the lateral diffusion in the slab. Substituting this expression into Eq. (4) leads to an important result – the profile of PTE depends on the aspect ratio of the waveguide $W_1/L$, see Fig. 3. It implies that in the slab geometry, the ratio between the impinging beam size and the thickness of the slab can be used to control the energy penetration. An analytical expression of the depth profile is given in SI, here, we only present the formula for the depth of the maximum of the intensity

$$z_{max} \simeq L/\left[1 + \sqrt{1 + 2L/W_1}\right]. \quad (9)$$

As expected, for $W_1 \gg L$ we recover known result $z_{max} \to L/2$. As $W_1/L$ decreases, the maximum of the energy profile is displaced towards the front surface. This result has practical applications as it offers a mechanism to scan the intensity maximum of PTE in the longitudinal direction of a disordered slab by varying the incident beam size.

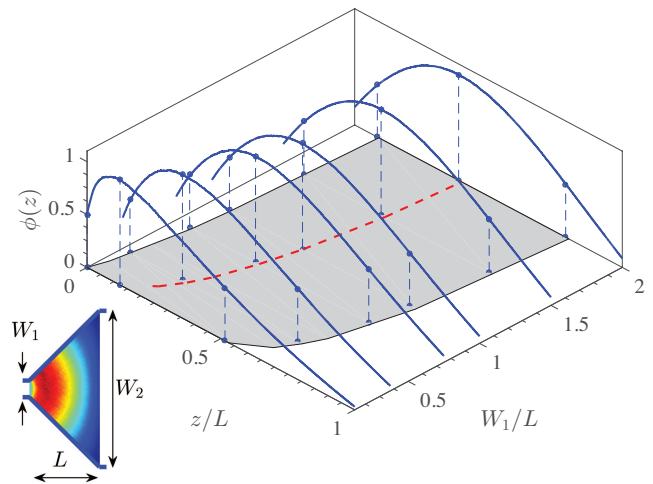

FIG. 3. (Color online) Prediction of PTE depth profile in an open slab with local illumination by approximating the lateral beam spreading with an expanding waveguide. The cross-section averaged intensity $\phi(z)$ for the linearly expanding waveguides $W(z) = W_1 + 2 \times z$ with different values of $W_1$ and fixed $L/\ell \simeq 18.3$. Gray area represents the region where $\phi(z)$ is greater than $1/2$ of its maximum value. The position of the maximum $z_{max}$ depends on $W_1/L$, which is well described by Eq. (9), bold dashed line.

In conclusion, we proposed a model for the perfectly transmitting eigenchannel (PTE) in the two-dimensional (2D) random system with an arbitrary shape. A self-action term was incorporated into the 2D diffusion equation for the return probability, to account for the interference effect. We employed a projection technique to reduce the 2D problem to 1D, and obtained an analytical expression relating the depth profile of the PTE to the boundary shape of the waveguide. This relation enabled the inverse design, namely, finding the waveguide shape to achieve the desired depth profile of the PTE. As an example, we predicted and verified numerically a specific shape of the waveguide in which PTE has a triangular profile. Such a profile, distinct from the universal parabolic-like profile of the PTE in the rectangle-shaped waveguide [4, 7, 9, 17], yields a tighter energy distribution, that can enhance the local light-matter interaction inside the diffusive medium.

Approximating the lateral beam spreading with an expanding waveguide, we predict the depth profile for the PTE in an open slab with local illumination. The depth for the maximum intensity increases with the size of the impinging beam. Our model can be further extended to include the effect of mismatched boundary conditions [32] (via an appropriate choice of $z_0$), and to describe three-dimensional geometries. In the latter case, the projection to one-dimension Eq. (2) shall be modified as well [20]. The results presented here are applicable to electromagnetic, acoustic, electronic and other types of waves. In optics, in particular, controllable delivery of light to different depths may lead to non-invasive imaging, sensing and therapeutic applications.

*Acknowledgments.* The authors acknowledge the support from the US National Science Foundation under Grant No. DMR-1205223, from the US Office of Naval Research under grant No. N00014-13-1-0649, and the US–Israel Binational Science Foundation (BSF) under grant No. 2015509.

# Supplementary Information:
# Inverse Design of Perfectly Transmitting Eigenchannels in Scattering Media

M. Koirala,[1] R. Sarma,[2] H. Cao,[2] and A. Yamilov[1, *]

[1]*Department of Physics, Missouri University of Science and Technology, Rolla, Missouri 65409,USA*
[2]*Department of Applied Physics, Yale University, New Haven, CT, 06520, USA*
(Dated: March 14, 2017)

PACS numbers: 42.25.Dd,42.25.Hz,73.23.-b

## A. NUMERICAL SIMULATIONS

We numerically obtain the perfectly transmitting eigenchannel (PTE) by the direct solution of the wave equation using KWANT simulation package [1, 2]. It allows to conveniently compute the transmission matrix $\hat{t}$ relating the incoming and outgoing wave amplitudes. The simulated system is a two-dimensional disordered waveguide $0 \leq z \leq L$, $|y| \leq W(z)/2$. In KWANT it is defined as a collection of coupled lattice sites in the two-dimensional rectangular grid described by a tight-binding Hamiltonian. Lack of bonds at the terminal sites at the sidewalls naturally introduces the reflecting boundary conditions. To model a passive random medium we introduce disorder by adding a random on-site potential $\delta E_{ii}$ to the diagonal elements as $H_{ii} = E_0 + \delta E_{ii}$, while keeping the nearest neighbor couplings at constant value of 1. The scattering region $0 \leq z \leq L$ is connected to the leads at $z < 0$ and $z > L$ where $\delta E_{ii} = 0$. This model is well suited to describe wave scattering phenomena as long as $k\ell \gg 1$ [3] where $k$ is the wave number and $\ell$ is transport mean free path.

The transmission matrix $\hat{t}$ relates the amplitudes of the propagating modes incident from the left lead $\phi_a$ to those of the outgoing modes in the right lead $\phi_b$. Representing $t^\dagger t = \hat{V}\hat{\tau}\hat{V}^\dagger$ gives the diagonal matrix of eigenvalues $\tau_n$ and the corresponding eigenchannels $V_{an}$. After computing $\hat{t}$, we construct the input field vector $\phi_a = V_{an}$ to couple exclusively into a specific eigenchannel $n$. With $\tau_n$ arranged in the decreasing order, $n = 1$ corresponds to the maximum transmission eigenchannel [2]. The parameters of the waveguides are chosen to be in the regime of a well-developed diffusive propagation $L/\ell, W/\ell \gg 1$. In this regime, the universal bimodal distribution of the eigenvalues $\tau_n$ yields the maximum at $\tau \simeq 1$. In each disorder realization we select eigenchannel with $n = 1$ and retain it only if $1 - \epsilon < \tau_{n=1} < 1$ with $\epsilon = 0.03$. Then we compute its intensity $I_{PTE}(\mathbf{r})$ and average over the ensemble of 1000 random realizations of disorder to obtain $\langle I_{PTE}(\mathbf{r}) \rangle$. $\phi(z)$ is obtained by averaging $\langle I_{PTE}(\mathbf{r}) \rangle$ over the cross-section of the waveguide.

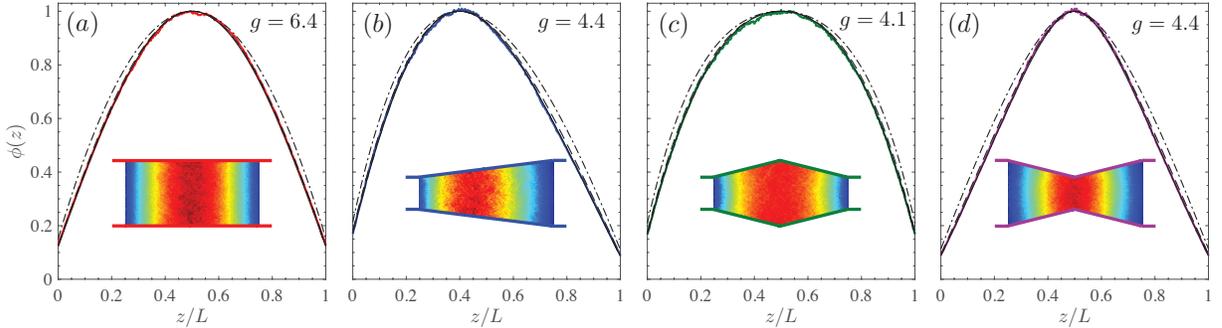

FIG. S1. (Color online) Demonstration of the importance of the self-action term in Eq. (3) of the main text. Panels show the cross-section averaged intensity $\phi(z)$ of the perfectly transmitting eigenchannel (PTE) in four waveguides of different shapes: (a) constant width (red), (b) linear expanding (blue), (c) lantern (green), (d) bow-tie (purple). $L/\ell \simeq 18.3$, the width at the narrowest point is $W_1/\ell \simeq 4.6$, the width at the widest point is $W_2/\ell \simeq 9.1$. The results of numerical simulations are plotted with colored bold solid lines. Solutions of Eq. (4) without and with the self-action term $\alpha$ are plotted with black dash-dotted and dashed lines respectively. Our model prediction agrees to the numerical result when the self action is included. A further improvement of the agreement is obtained by accounting for the localization-induced correction of the diffusion coefficient (thin solid lines) in these systems with relatively small values of dimensionless conductance $g$, quoted in each panel. A quantitative evaluation of the agreement is given in Table I.

## B. SELF-ACTION TERM

In this section, we illustrate the importance of the self-action term in Eq. (2,3) by comparing the depth profile of PTEs with and without this term (the second term proportional to $\alpha$). In other words, we test the importance of $\alpha \neq 0$ in our model. Fig. S1 shows the depth profile $\phi(z)$ computed numerically (solid bold lines) and with the following three approximations:
(i) Solution of Eqs. (2,3) with $\alpha \equiv 0$, and no localization corrections to the diffusion coefficient, i.e. $D(z) = D_0$, – dash-dotted lines;
(ii) Solution of Eqs. (2,3) with $\alpha$ chosen as discussed in the main text, and $D(z) = D_0$, – dashed lines;
(iii) Solution of Eqs. (2,3) with $\alpha$ chosen as discussed in the main text, and the position-dependent diffusion coefficient $D(z)/D_0 = [1 + F(z)/(2g)]^{-1}$ found with the self-consistent theory of localization [4] – thin solid lines. The return probability function $F(z)$ was calculated from Eq. (4) in the main text. The dimensionless conductance $g$ was obtained from the numerical simulation, its value is quoted in each panel in Fig. S1. To quantify the discrepancies between theory and simulations we define the deviation between two intensity profiles $I_{1,2}(z)$ as

$$\Delta = \left[ \frac{L \int_0^L |I_1(z) - I_2(z)|^2 dz}{\int_0^L I_1(z)dz \int_0^L I_2(z)dz} \right]^{1/2}. \quad (S1)$$

|  | Rectangular | Horn | Lantern | Bowtie |
|---|---|---|---|---|
| $\alpha = 0$ | 5.7% | 6.8% | 5.8% | 7.3% |
| $\alpha \neq 0$ | 1.0% | 1.9% | 1.7% | 1.7% |
| $\alpha \neq 0$ & $D(z)$ | 0.9% | 0.9% | 0.8% | 1.0% |

TABLE I. Deviation in the depth profile of PTE between numerical simulations and model predictions in Fig. S1, described by Eq. (S1). First row "$\alpha = 0$" – the deviation between dash-dotted black lines (Eq.(4) without self-action term) and bold colored lines (numerical result); "$\alpha \neq 0$" – the deviation between dashed black lines (Eq.(4) with self-action term) and bold colored lines; "$\alpha \neq 0$ & $D(z)$" – the deviation between thin solid black lines (Eq.(4) with self-action term and localization correction of diffusion coefficient) and bold colored lines. Four columns correspond to the four waveguide shapes shown in Fig. S1(a-d).

The numerical value of the deviation for each curve in Fig. S1 is listed in Table I.

We find that accounting for the self-action term, that is proportional to $\alpha$ in Eq. (2), leads to agreement between our model prediction and numerical simulations. Due to relatively small values of the dimensionless conductance $g$, the localization-induced correction of the diffusion coefficient improves the agreement even further. The significance of this correction, however, is diminished even for systems with a moderately large values of $g$.

## C. FUNDAMENTAL MODE OF DIFFUSION EQUATION

In Fig. 1 of the main text we confirmed previous reports that the model based on the fundamental mode (FM) of the diffusion equation gives a good prediction for the depth profile of PTE in the constant-width waveguide [5, 6]. Here, we further analyzes the applicability of this model to waveguides of shapes other than rectangle.

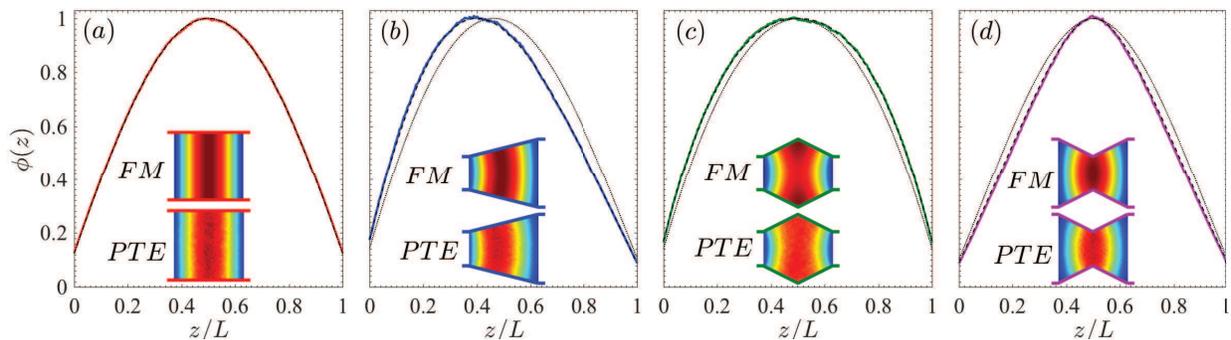

FIG. S2. (Color online) Test of FM-model in diffusive waveguides of different shapes. (a-d) plot the depth profile of PTE in four geometries shown in insets. The results of numerical simulations are shown with colored bold solid lines. Predictions of the FM-model are plotted by dotted lines, whereas those of our model, Eq. (4), are shown as dashed lines. The corresponding 2D energy density distribution for the fundamental mode ("FM") and the perfectly transmitting eigenchannel ("PTE") are shown as insets. System parameters are identical to Figs. 1(a,b) of the main text. The quantitative comparison via Eq. (S1) of the deviations between the model predictions and the numerical simulations is given in Table II.

Figure S2 presents the results of numerical simulations (colored bold solid lines) and compares them to the predictions of FM-model (black dotted lines) and to that of our model Eq. (4). System parameters ($\ell$, $W$'s, $L$) are the same as in Figs. 1(a,b). Due to the relatively larger value of dimensionless conductance ($g \sim 10$) in these systems, the localization corrections are further reduced and, therefore, were not included in this calculation. We see

|        | Rectangular | Horn  | Lantern | Bowtie |
|--------|-------------|-------|---------|--------|
| FM     | 0.4%        | 10.7% | 5.8%    | 7.0%   |
| Eq.(4) | 0.4%        | 1.0%  | 0.7%    | 1.9%   |

TABLE II. Deviation in the depth profile of PTE between numerical simulations and model predictions in Fig. S2, described by Eq. (S1). "FM" (the previous model based on the fundamental mode of diffusion equation) – the deviation between dotted black lines and bold colored lines; "Eq. (4)" (our model expressed by Eq. (4) in the main text) – the deviation between dashed black lines and bold colored lines. Four columns correspond to the four waveguide shapes in Fig. S2(a-d).

that FM-model does not predict the depth profile in systems with non-constant width.

Further insight can be gained by comparing the full two-dimensional structure of the fundamental mode (FM) of the 2D diffusion equation (upper inset annotated as "FM") to that of the numerically calculated PTE (the lower insets). We note that the FM model was developed for 1D diffusion in a slab geometry, and its application to other geometries was not known. Our results show that in most geometries there exist a qualitative agreement between FM and PTE, but in the lantern geometry shown in panel (c) we observe a systematic qualitative deviation, see Table II. Indeed, while the maximum intensity of FM occurs in the two corners at $z = L/2$, PTE peaks at the center of the cross section at $z = L/2$. This difference can be attributed to the difference in construction of the two functions. The FM is the eigenmode of the diffusion equation with the smallest decay constant, i.e. the lowest leakage through open boundaries. Therefore, the maximum of stored energy is at the locations further away from the two ends of the waveguide, which are the corners at $z = L/2$ in panel (c). In contrast, PTE maximizes the transmission through the system and, therefore, concentration of energy along the "line-of-sight" from the input end to the output end is preferable. Since no inner corners exist in the rectangle geometry, FM coincides with PTE.

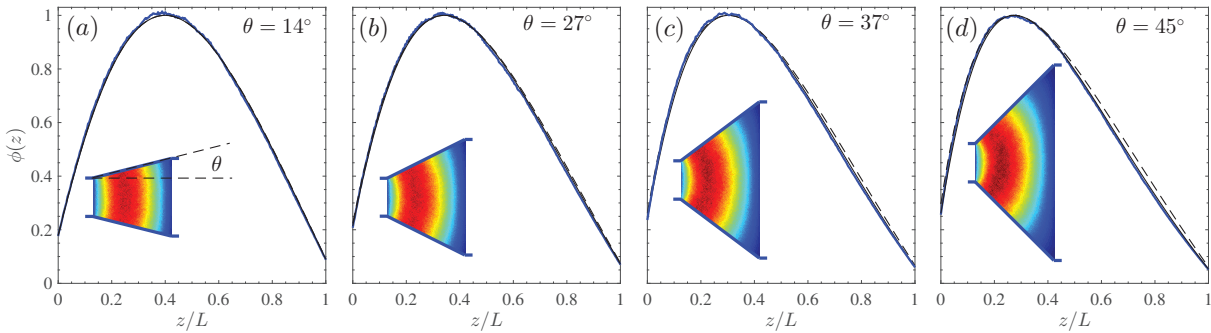

FIG. S3. (Color online) Test of the projection method. Four panels correspond to an increasing angle $\theta$, defined as $\tan\theta = W'(z)/2$, see panel (a), and quoted in the upper right corner of each panel. Each panel plots the PTE profile computed (i) numerically (colored bold solid line), (ii) by solving Eq. (1) in the main text (black thin solid line), and (iii) by solving Eq. (3) with projection Eq. (2) in the main tex (black thin dashed line). In all cases $L/\ell \simeq 18.3$ and $W_1 = L/2$. The quantitative comparison via Eq. (S1) of the deviations between the models and numerical simulations is summarized in Table III.

### D. TEST OF PROJECTION ANSATZ

Reduction of the two-dimensional problem in Eq. (1) to the one-dimensional problem in Eq. (2) of the main text relied on the projection technique via Eq. (3). It is expected (see e.g. Ref. [7] for a review and the detailed citations) that such an approach remains accurate for the tapering angle below 45°, i.e., $W'(z)/2 \leq 1$. It is important to check the validity of the projection method in our case, because the data reported in Fig. 3 of the main text corresponds to the borderline case $W'(z)/2 = 1$. To this end, we performed calculations of $\phi(z)$ using two methods: numerically solving (i) the original two-dimensional Eq. (1), and (ii) the reduced one-dimensional version in Eq. (2) that employs the projection ansatz in Eq. (3). The results of our simulations for the increasing values of $W'(z)/2 = 1/4, 1/2, 3/4, 1$ are shown in Fig. S3. These four cases correspond to the tapering angle of waveguide boundary $\theta = 14°, 27°, 37°, 45°$. Bold blue solid lines represent numerical simulations, black thin solid lines – 2D model, black thin dashed lines – projection to 1D. We find that even at very large angle of $\theta = 45°$ the deviation is rather small $< 4\%$, see Table III. Furthermore, we see that the original 2D model describes the numerical results better, with deviations $< 2\%$. We conclude that our estimation of beam spreading in the main text, in particular, Eq. (9) remains applicable, which is further confirmed in Fig. 3 of the main text showing accurate prediction of the depth of the maximum of the energy density $z_{max}$.

| $\theta$ | 14° | 27° | 37° | 45° |
|---|---|---|---|---|
| Projection ansatz | 1.2% | 2.3% | 2.9% | 3.9% |
| 2D model | 1.1% | 1.6% | 1.8% | 1.8% |

TABLE III. Deviation in the depth profile $\phi(z)$ of PTE between numerical simulations and model calculations in Fig. S3, described by Eq. (S1). "Projection ansatz" – Eq. (4): the deviation between dashed black lines and bold colored lines; "2D model" – Eq. (1): the deviation between black solid lines and bold colored lines. Four columns correspond to the four waveguide shapes in Fig. S3(a-d).

### E. PTE PROFILES FOR SELECT GEOMETRIES

In the main text we outlined the procedure to find the depth profile of PTE $\phi(z)$ for a given shape of disordered waveguide $w(z)$, Eqs. (2,4), or other way around, to find $w(z)$ from $\phi(z)$, Eq. (6). In this section we present the pairs $w(z)$, $\phi(z)$ for a select group of waveguide geometries studied in this work.

1. Rectangular waveguide:

$$w(z) \equiv 1$$
$$\phi(z) = \frac{4(1-\tilde{\alpha})\zeta(1-\zeta)}{1 - 4\tilde{\alpha}\zeta(1-\zeta)}$$
$$\zeta = (z+z_0)/(L+2z_0).$$

2. Linearly expanding waveguide:

$$w(z) = \left(\frac{W_1}{W_2 - W_1} + \frac{z}{L}\right) \log \frac{W_2}{W_1}$$
$$\phi(z) = \frac{4(1-\tilde{\alpha})\zeta(1-\zeta)}{1 - 4\tilde{\alpha}\zeta(1-\zeta)}$$
$$\zeta = \frac{\dfrac{z_0}{W_1} + \dfrac{L}{W_2 - W_1} \log \dfrac{w(z)}{w(0)}}{\dfrac{z_0}{W_1} + \dfrac{z_0}{W_2} + \dfrac{L}{W_2 - W_1} \log \dfrac{w(L)}{w(0)}}.$$

Here $W_{1,2}$ denote the widths at $z = 0, L$.

3. Bowtie/lantern waveguide:

$$w(z) = \left(\frac{W_2}{W_1 - W_2} + \left|2\frac{z}{L} - 1\right|\right) \log \frac{W_1}{W_2}$$
$$\phi(z) = \frac{4(1-\tilde{\alpha})\zeta(1-\zeta)}{1 - 4\tilde{\alpha}\zeta(1-\zeta)}$$
$$\zeta = \frac{\dfrac{z_0}{W_1} + \dfrac{L/2}{W_2 - W_1} \log \dfrac{w(z)}{w(0)}}{\dfrac{z_0}{W_1} + \dfrac{z_0}{W_2} + \dfrac{L}{W_2 - W_1} \log \dfrac{w(L)}{w(0)}}.$$

Here $W_1$ denote the widths at $z = 0, L$, and $W_2$ is the width at the mid-point $z = L/2$ of the waveguide.

---